\documentclass[global,twocolumn]{svjour}

\usepackage{graphicx, color}
\usepackage{dcolumn}
\usepackage{bm}
\usepackage{booktabs}

\begin{document}

\title{Fictitious Magnetic Resonance by Quasi-Electrostatic Field}

\author{Jun Kobayashi\inst{1} \and Kosuke Shibata\inst{1} \and Takatoshi Aoki\inst{1} \and Mitsutaka Kumakura\inst{1,2,3} \and Yoshiro Takahashi\inst{1,2}}
\institute{Department of Physics, Graduate School of Science, Kyoto University, Kyoto 606-8502, Japan
\and
CREST, JST, 4-1-8 Honcho Kawaguchi, Saitama 332-0012, Japan
\and
PREST, JST, 4-1-8 Honcho Kawaguchi, Saitama 332-0012, Japan
}

\date{\today}
\maketitle
\begin{abstract}
We propose a new kind of spin manipulation method using a {\it fictitious} magnetic field generated by a quasi-electrostatic field. 
The method can be applicable to every atom with electron spins and has distinct advantages of small photon scattering rate and local addressability. 
By using a $\rm{CO_2}$ laser as a quasi-electrostatic field, we have experimentally demonstrated the proposed method by observing the Rabi-oscillation of the ground state hyperfine spin $F=1$ of the cold $\rm{^{87}Rb}$ atoms and the Bose-Einstein condensate.

\end{abstract}

\section{Introduction}
A spin of atom is a fundamental observable in physics, and the coherent manipulation of a spin is also very important. 
There are two well-known methods of manipulating atomic spins by applying a resonant external field, magnetic resonance with a radio-frequency (RF) magnetic field\cite{JJsakurai} and stimulated Raman transitions induced by near resonant electromagnetic fields\cite{suter}.
These techniques have been successfully applied in many studies of science. 
In particular, they are useful for coherently manipulating qubits of atomic spins in quantum information processing\cite{entanglement}. 

In this paper, differently from the above methods, we propose a new method for coherently manipulating atomic spins using a quasi-electrostatic field. 
The advantages of the new method are that it results in a very small decoherence, because the photon scattering rate induced by the quasi-electrostatic field is negligibly small compared to that of usual stimulated Raman transitions, and that it has better local addressability than the conventional magnetic resonance. 

Since the deep potential depth can be achieved by the tightly focused high power beam with negligible photon scattering rate, the quasi-electrostatic field has been used in many experiments for optical trapping of cold atoms\cite{Takekoshi,haensch} and molecules\cite{Cs molecule Takekoshi,Rb molecule}, and has been successfully used in the formation of Bose-Einstein condensate (BEC) and Fermi degenerate state\cite{Chapman,Li atom Thomas,Cs,APB}. 
The trap potential, or light shift for quasi-electrostatic fields is well described by using the polarizability of the static electric field, for which the interaction is completely independent of a spin. 
Therefore, the spin-dependence of the light shift for the quasi-electrostatic field has escaped attention so far. 
However, our careful analysis has shown that the spin-dependence of the light shift for the tightly focused high power laser of quasi- electrostatic field is by no means negligibly small and in fact large enough to induce the magnetic resonance of the atomic spin, which has been successfully demonstrated in the present experiment.

For near resonant fields, the spin-dependence of the light shift is widely used in many experiments and can be interpreted in terms of a fictitious magnetic field\cite{Cohen-Tannoudji}.
In the pioneering experiment by Cohen-Tannoudji and Dupont-Roc, the Zeeman shift due to the fictitious magnetic field was observed and a magnetic resonance induced by modulating the fictitious magnetic field was demonstrated\cite{Cohen-Tannoudji}. 
The fictitious magnetic field has recently been also used for observation of the spin precession and the spin echo\cite{ficecho1,ficecho2}. 
These techniques which used near resonant fields, however, are not straightforwardly applicable to optically trapped cold atoms because of the heating effect by the photon scattering. 
By using the quasi-electrostatic field, we can overcome this difficulty and make use of the novel techniques using the fictitious magnetic field for the cold atoms and BEC.

Our experimental demonstration of the magnetic resonance using the fictitious magnetic field due to a quasi-electrostatic field has been done by observing the Rabi oscillation of the spin of cold $\rm{^{87}Rb}$ atoms and BEC by using a $\rm{CO_2}$ laser as a quasi-electrostatic field. 
Our experiment is also the first quantitative measurement of the interaction between a quasi-electrostatic field and an electron spin. 
In the following, we describe in detail the principle of our method and its experimental demonstration. 

\section{Fictitious Magnetic Resonance}
First we explain the fictitious magnetic field induced by a laser field of the frequency $\omega_L$ for alkali atoms\cite{Grimm}. 
As long as the detuning is large compared to the hyperfine splitting of the excited state, the hyperfine splitting is neglected and the light shift can be written as the sum of the contributions of the $D_1$ and $D_2$ transitions:
\begin{eqnarray}
\label{eq:1}
U_q&=&-\frac{d^2}{6{\epsilon_0}{\hbar}c}\left(\frac{2+g_Fm_Fq}{\omega_{D_2}-\omega_L}+\frac{1-g_Fm_Fq}{\omega_{D_1}-\omega_L}\right)I.
\end{eqnarray}
Here $d$ is the dipole moment defined as $d\equiv|{\langle}J=1/2||er||J^{\prime}=1/2{\rangle}|$ ($J$ and $J^{\prime}$ are the total electron angular momentum of the ground and the excited states, respectively.) $\epsilon_0$, $\hbar$, and $c$ are the dielectric constant, Planck's constant, and velocity of light, respectively. 
$\omega_{D_1}$ and $\omega_{D_2}$ are the resonant frequencies of the $D_1$ and $D_2$ transitions, respectively. $m_F$ and $g_F$ are the magnetic quantum number and the Land$\rm{\acute{e}}$ g-factor of the ground state, respectively.  $I$ is the intensity of the laser. $q$ represents the polarization of the laser ($q=0, \pm1$ for $\pi$, $\sigma^{\pm}$ polarization light). The $m_F$ dependence of the light shift can be understood as the effect that the atom feels a fictitious magnetic field, since the $m_F$ dependence is the same as the Zeeman shift for a weak magnetic field \cite{Cohen-Tannoudji}.
The light shift can be rewritten by using the fictitious magnetic field, which is defined as follows:
\begin{eqnarray}
\label{eq:2}
U_q&=&U_0+g_Fm_F{\mu}_BB_{\rm{fic}},\\
\label{eq:3}U_0&\equiv&-\frac{d^2}{6{\epsilon_0}{\hbar}c}\left(\frac{2}{\omega_{D_2}-\omega_L}+\frac{1}{\omega_{D_1}-\omega_L}\right)I,\\
\label{eq:4}{\mu}_BB_{\rm{fic}}&\equiv&\frac{d^2}{6{\epsilon_0}{\hbar}c}\frac{\omega_{D_2}-\omega_{D_1}}{(\omega_{D_2}-\omega_L)(\omega_{D_1}-\omega_L)}qI.
\end{eqnarray}
Here $U_0$ is spin-independent part of the light shift and corresponds to the light shift for $\pi$ polarization laser. 
${\mu}_B$ is Bohr magneton. 
The direction of the fictitious magnetic field is parallel to the direction of the wave vector.
As shown in Eq. (\ref{eq:4}), $B_{\rm{fic}}$ is proportional to $\omega_{D_2}-\omega_{D_1}$, which means that the electron spin in the ground state has the indirect interaction with the electric field through the spin-orbit coupling in the excited state. 

For the quasi-electrostatic field ($\omega_L\ll\omega_{D_1}, \omega_{D_2}$), the counter rotating term with the frequency $-\omega_L$, which is ignored in the usual rotational wave approximation, must be considered. 
Here we must keep in mind that the polarization of the $-\omega_L$ field is opposite to that of the $\omega_L$ field. 
Therefore $B_{\rm{fic}}$ for a quasi-electrostatic field can be written as the sum of the contributions of $(\omega_L, q)$ and $(-\omega_L, -q)$:
\begin{eqnarray}
\label{eq:5}
{\mu}_BB_{\rm{fic}}&=&\frac{d^2}{6{\epsilon_0}{\hbar}c}\frac{\omega_{D_2}-\omega_{D_1}}{(\omega_{D_2}-\omega_L)(\omega_{D_1}-\omega_L)}qI\nonumber\\
&+&\frac{d^2}{6{\epsilon_0}{\hbar}c}\frac{\omega_{D_2}-\omega_{D_1}}{(\omega_{D_2}+\omega_L)(\omega_{D_1}+\omega_L)}(-q)I\nonumber\\
&\approx&\frac{d^2}{3{\epsilon_0}{\hbar}c}\frac{\omega_{\rm{D_2}}^2-\omega_{\rm{D_1}}^2}{\omega_{\rm{D_1}}^2\omega_{\rm{D_2}}^2}\omega_{\rm{L}}qI.
\end{eqnarray}
Here the approximation of $\omega_L\ll\omega_{D_1}, \omega_{D_2}$ is used in the last equation. In order to obtain general expression, we should sum the contributions from all excited $P_{1/2}$ and $P_{3/2}$ states. However, the higher excited state has the smaller fine-structure splitting and has smaller transition moments with the ground state. Then the contributions from the lowest two levels of $P_{1/2}$ and $P_{3/2}$ are dominant. Therefore, we ignore the contributions from the excited states except for the lowest two levels in this paper\cite{Footnote1}.

\begin{figure}
\begin{center}
\resizebox{8.0cm}{!}{\includegraphics{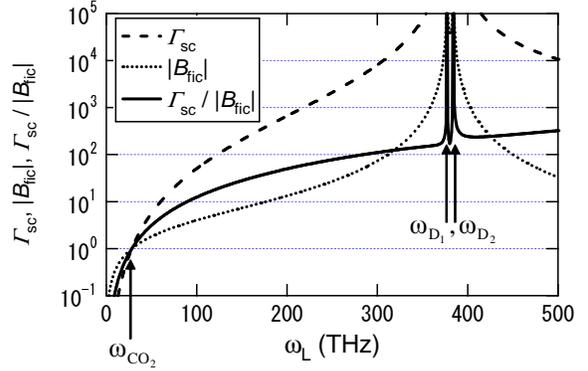}}
\caption{\label{fig:1} Fictitious magnetic field and photon scattering rate for Rb atom. 
Fictitious magnetic field ($|B_{\rm{fic}}|$), photon scattering rate ($\Gamma_{\rm{sc}}$), and their ratio ($\Gamma_{\rm{sc}}/|B_{\rm{fic}}|$) are plotted as a function of the laser frequency. 
Here these values are normalized by the values at the $\rm{CO_2}$ laser frequency. 
$\Gamma_{\rm{sc}}/|B_{\rm{fic}}|$ at the $\rm{CO_2}$ laser frequency is about 100 times smaller than that at the near resonant frequency.
}
\end{center}
\end{figure}

Figure \ref{fig:1} shows the fictitious magnetic field, the photon scattering rate, and the ratio $\Gamma_{\rm{sc}}/|B_{\rm{fic}}|$ for Rb atom as a function of the laser frequency.
Here the values are normalized by the values at the $\rm{CO_2}$ laser frequency ($2\pi\times2.83\times10^{13}$ Hz) which is much smaller than the optical transition frequencies $\omega_{\rm{D_1}}$ ($2\pi\times3.77\times10^{14}$ Hz ) and $\omega_{\rm{D_2}}$ ($2\pi\times3.85\times10^{14}$ Hz ) of the $\rm{^{87}Rb}$ atoms.
The photon scattering rate is also written as the sum of the contributions of $D_1$ and $D_2$ transitions:
\begin{eqnarray}
\label{eq:6}
\Gamma_{\rm{sc}}=\frac{2d^4}{9{\epsilon_0}^2{\hbar}^3c^4}\left[\frac{\omega_{\rm{D_1}}^2}{(\omega_{\rm{D_1}}^2-\omega_{\rm{L}}^2)^2}+\frac{2{\omega_{\rm{D_2}}^2}}{(\omega_{\rm{D_2}}^2-\omega_{\rm{L}}^2)^2}\right]\omega_{\rm{L}}^3I.
\end{eqnarray}
As shown in Eq. (\ref{eq:6}), $\Gamma_{\rm{sc}}$ is proportional to $\omega_{\rm{L}}^3$, in contrast to the fact that the $B_{\rm{fic}}$ is proportional to $\omega_{\rm{L}}$ as shown in Eq. (\ref{eq:5}), and thus $\Gamma_{\rm{sc}}$ is more strongly suppressed than $B_{\rm{fic}}$ in the low frequency region. 
The benefit of using a quasi-electrostatic field is clear from the fact that $\Gamma_{\rm{sc}}/|B_{\rm{fic}}|$ at the $\rm{CO_2}$ laser frequency is about 100 times smaller than that at the near-resonant frequency as shown in Fig. \ref{fig:1}.  

\begin{figure}
\resizebox{8.0cm}{!}{\includegraphics{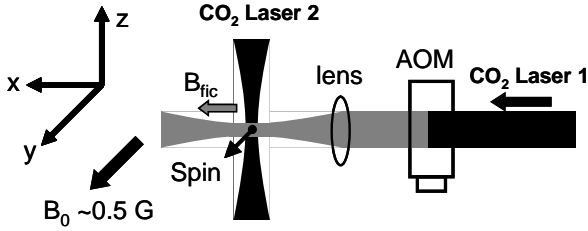}}
\caption{\label{fig:2} Experimental setup. The Rb atoms are trapped at the crossed region of the focused $\rm{CO_2}$ lasers 1 and 2. The static magnetic field ($B_0$) is perpendicular to the directions of both $\rm{CO_2}$ lasers. When the intensity of the $\rm{CO_2}$ laser 1 is modulated by an acousto-optic modulator (AOM) at the frequency of the Zeeman splitting of atom, the Rabi oscillation of the atomic spin is induced.}
\end{figure}

\section{Experimental}
We have performed an experiment to demonstrate the proposed method using $\rm{^{87}Rb}$ atoms loaded to the optical trap formed by the crossed $\rm{CO_2}$ lasers as shown in Fig. \ref{fig:2}. 
The detailed scheme of the cooling and loading of atoms is reported in Ref. \cite{APB}.
As a quasi-electrostatic field to generate a fictitious magnetic field, one of the $\rm{CO_2}$ lasers used for trapping is utilized ($\rm{CO_2}$ laser 1). 
The static magnetic field of about 0.5 G has been applied along the direction perpendicular to the propagation directions of two $\rm{CO_2}$ lasers, and the atoms are optically pumped to the $F=1, m_F=-1$ state. 
Then the intensity modulation is applied with an acousto-optic modulator (AOM) for one of the trapping lasers. 
When the modulation frequency is resonant to the Zeeman splitting of the $F=1$ atoms, the Rabi oscillation is induced. 
Each spin state of $F=1, m_F=0, \pm1$ is selectively observed by two-step imaging scheme. 
First the atoms in the $F=1$ hyperfine state are spin-selectively transferred to the $F=2$ hyperfine state by the usual stimulated Raman transition by virtue of the different Zeeman shifts, and then the atoms in the $F=2$ hyperfine state are detected by the absorption imaging method with $5^2S_{1/2},F=2{\rightarrow}5^2P_{3/2},F=3$ transition. 

\begin{figure}
\resizebox{8.0cm}{!}{\includegraphics{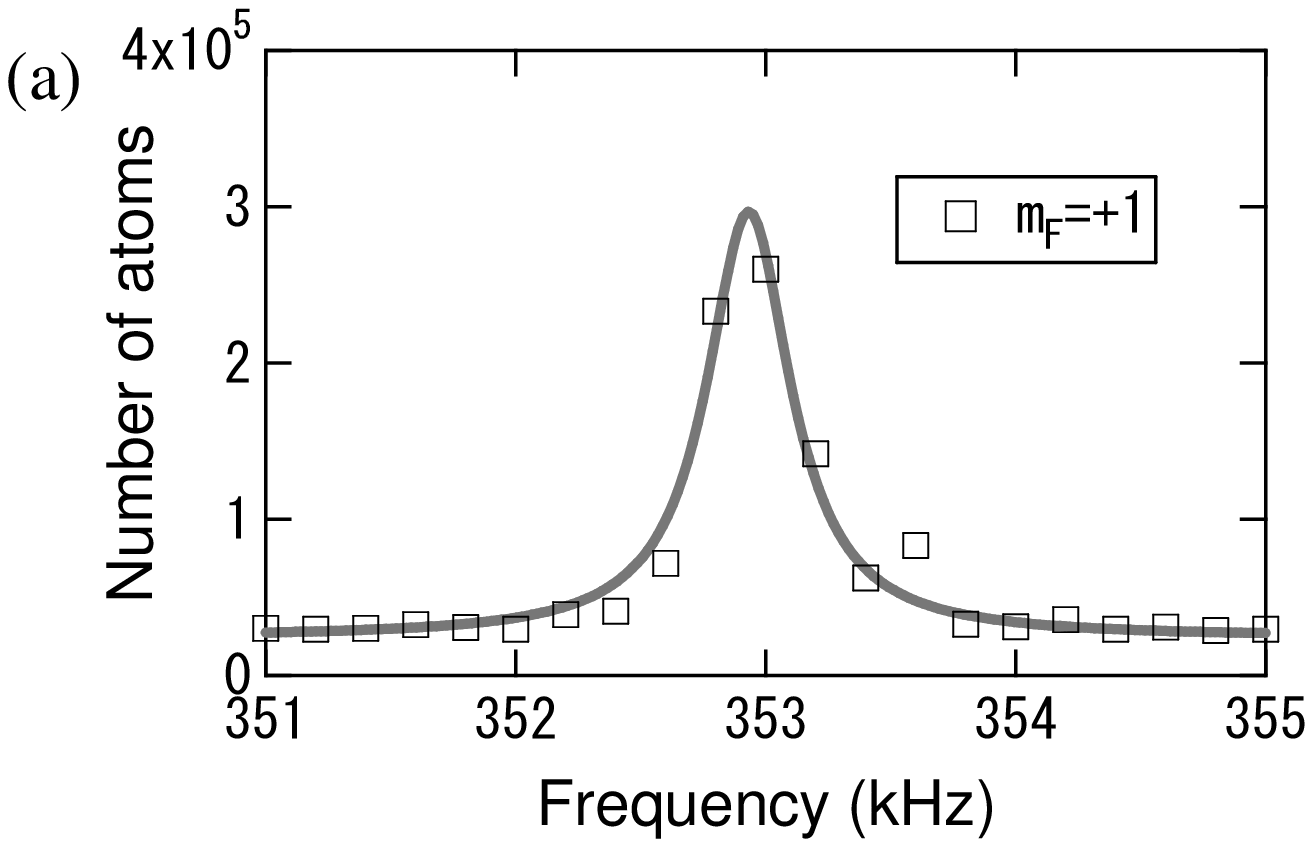}}
\resizebox{8.0cm}{!}{\includegraphics{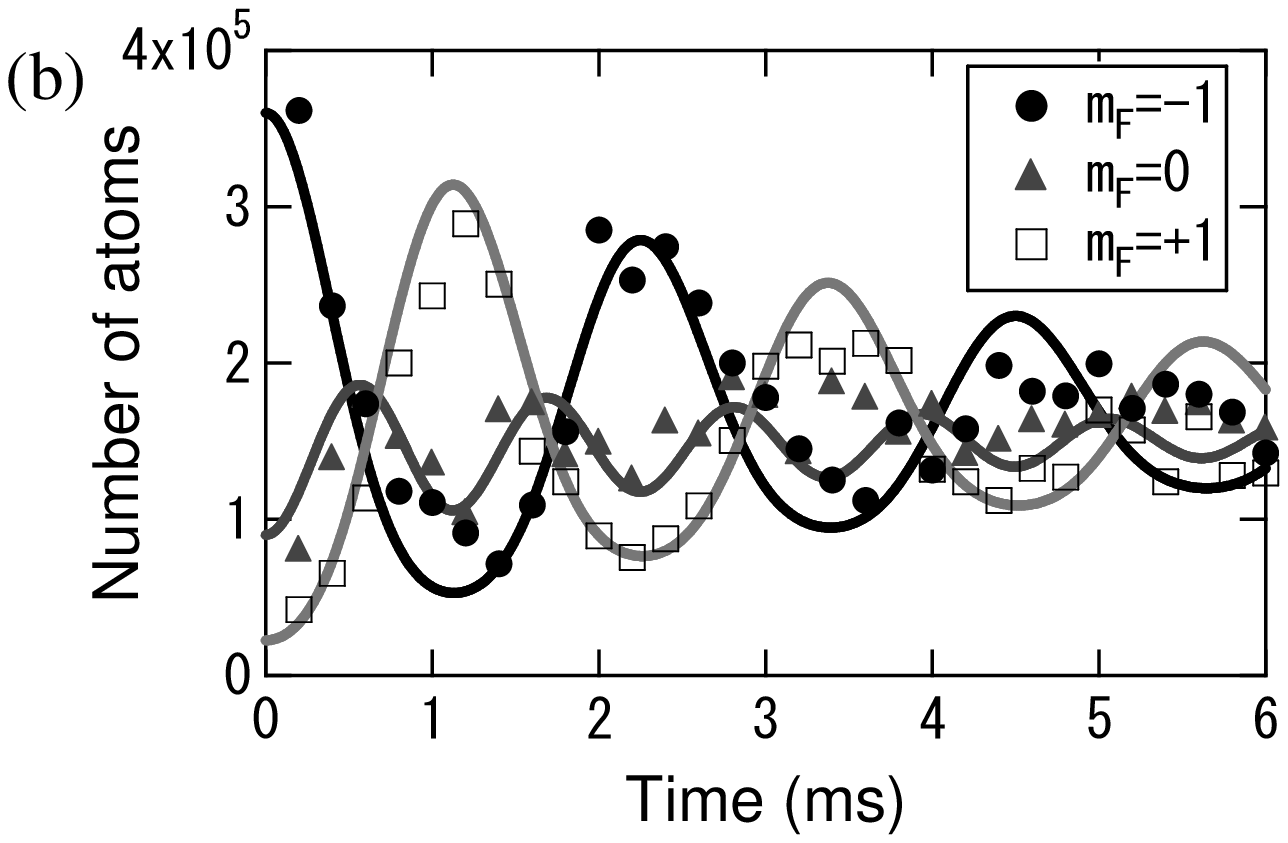}}
\resizebox{8.0cm}{!}{\includegraphics{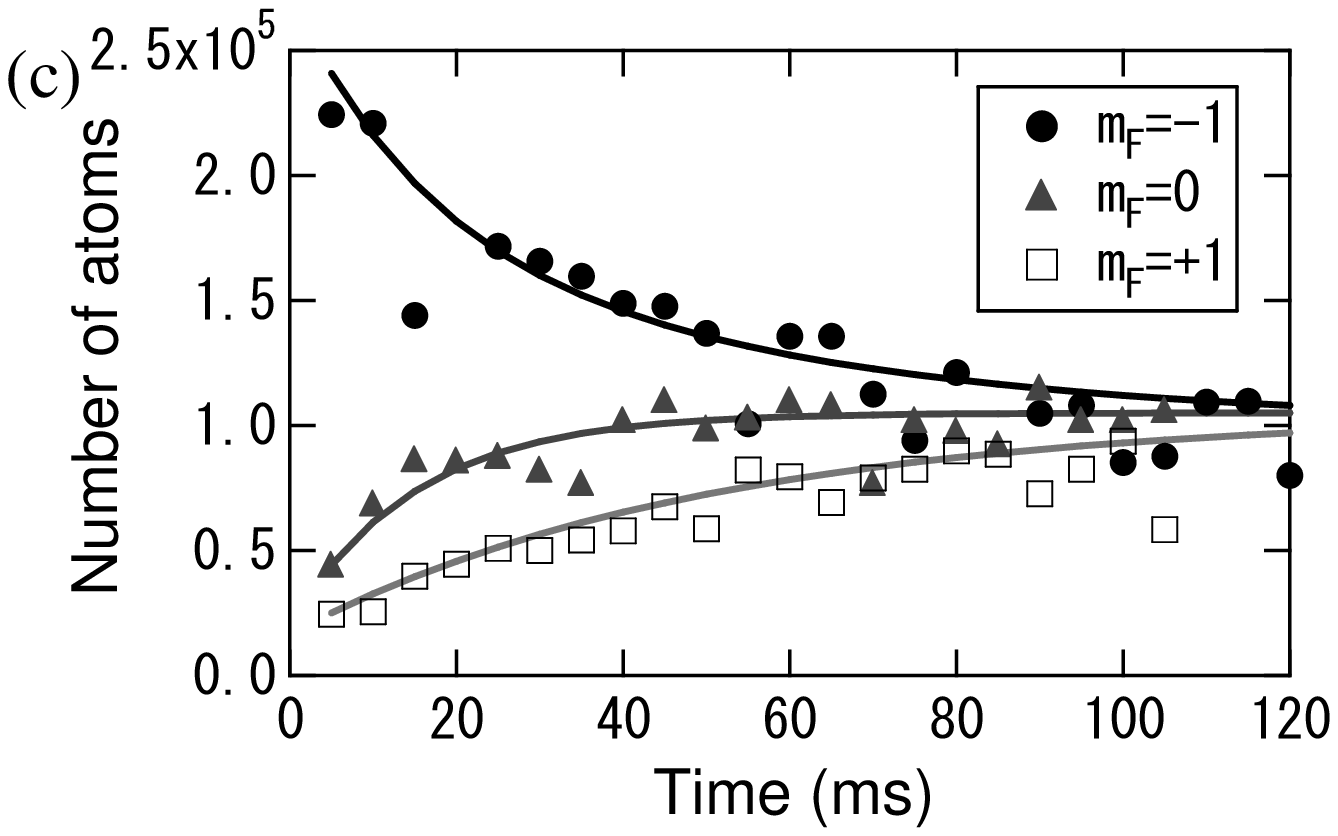}}
\caption{\label{fig:3}(a) Spectrum of the magnetic resonance induced by the quasi-electrostatic fields with circular polarization. The solid curve is the fitting using Lorentzian function. The linewidth is about 0.4 kHz. (b) Rabi oscillation between magnetic sublevels at the resonant frequency. The three components of $m_F=0, \pm1$ oscillate at the same frequency, and relax to the mean value of them. The solid curves are the numerical calculation with the Rabi frequency of 0.45 kHz and relaxation time of 4.5ms. (c) Spin relaxation when the linearly polarized $\rm{CO_2}$ laser is used. The solid curves are fitting to the data. This result means that the Rabi frequency for linearly polarized laser is much smaller than that for circularly polarized laser. This resonance is thought to be induced by a possible small portion of circular polarization component of $\rm{CO_2}$ laser.}
\end{figure}

The incident power of the $\rm{CO_2}$ laser is about 20 W and the beam waist at the focus is 62 $\rm{{\mu}m}$.
Then the peak intensity $I$ is about $3.3\times10^9$ $\rm{W/m^2}$. Using Eq. (\ref{eq:5}), the energy shift by the $B_{\rm{fic}}$ is calculated to be $2\pi\times2.2$ kHz ($B_{\rm{fic}}=3.3$ mG.) for the circularly polarized light ($q=\pm1$). The Rabi frequency induced by the field is estimated to be $2\pi\times550$ Hz, which is large enough to be experimentally observed, though the energy shift by the $B_{\rm{fic}}$ is about 2000 times smaller than the spin independent light shift $U_0$. 

The experimental results for circularly polarized light are shown in Fig. \ref{fig:3}(a) and (b). Figure \ref{fig:3}(a) shows the obtained resonance spectrum, where the modulation time is 1 ms. The observed resonant frequency of about 353 kHz is consistent with the magnetic field value of 0.5 G measured by the stimulated Raman transition spectrum. The observed linewidth of about 0.4 kHz is power broadened. Figure \ref{fig:3}(b) shows the time evolutions of the spin components $m_F=0, \pm1$ when the modulation frequency is resonant. One can see the clear Rabi oscillation between the magnetic sublevels, which is the demonstration of the coherent manipulation of the atomic spins with our method. The observed Rabi frequency is about $2\pi\times450$ Hz, which is a little smaller than the expected value. We think that the intensity modulation and the polarization of the $\rm{CO_2}$ laser are not ideal, which we could not measure in good accuracy. The relaxation time of the Rabi oscillation is about 4.5 ms, which may be caused by the possible stray magnetic field or inhomogeneity of the intensity of the $\rm{CO_2}$ laser, which cannot be avoided in our present experimental setup.

The theoretical analysis given in the above (Eq. (\ref{eq:5})) suggests the absence of the fictitious magnetic field for linearly polarized light ($q=0$). Figure \ref{fig:3}(c) shows the experimental results for the linearly polarized light, where the spin components do not oscillate but slowly relaxes to the mean value of three components. This result means that the Rabi frequency for linearly polarized light is much smaller than that for circularly polarized light. We think that a possible small portion of circular polarization component of the $\rm{CO_2}$ laser induced the observed transition.

We have also applied the method for the spinor BEC of Rb atoms. 
In this case, the spin components can be spatially divided by applying magnetic field gradient after the atoms are released from the optical trap.
For the intensity modulation with the modulation frequency resonant to the Zeeman splitting of the $F=1$ state, we have clearly observed all the three $m_F=1, 0, -1$ spin components of BEC, otherwise only the $m_F=-1$ component of BEC was observed, which indicates that the spin manipulation can be successfully induced for BEC with no heating of atoms by quasi-electrostatic field.

This new method similarly works for other alkali-atoms and also for many atoms with a spin. The strength of the fictitious magnetic field for quasi-electrostatic field can be calculated by the known atomic parameters. Table \ref{tbl:1} shows $B_{\rm{fic}}$ for some atoms in comparison with Rb atom. For alkali atoms $B_{\rm{fic}}$ is proportional to the spin-orbit coupling strength as is given in Eq. (\ref{eq:5}), thus for heavier alkali atoms, $B_{\rm{fic}}$ is larger. For Group II atoms and Yb atom, the ground state has no electron spin but only nuclear spin for fermionic isotopes. The interaction between the nuclear spin and the electric field is only possible through the hyperfine coupling in the excited state. Thus the value of $B_{\rm{fic}}$ is much smaller than those of alkali atoms. Note that we define $B_{\rm{fic}}$ as in Eq. (\ref{eq:4}) using Bohr magneton ${\mu}_B$, instead of nuclear Bohr magneton. In contrast, for the atomic state with orbital angular momentum such as the ground $P$ state of Tl, $B_{\rm{fic}}$ is larger than that of most of the alkali atoms, since the orbital angular momentum interacts with the electric field directly. 

\begin{table}
\begin{center}
\caption{\label{tbl:1} Fictitious magnetic field induced by a quasi-electrostatic field. For alkali atoms, the fictitious magnetic field is larger for heavier atoms. For atoms with no electron spins, it is very small. In contrast, for the atomic state with orbital angular momentum, it is larger than that of most of the alkali atoms. 
}
\begin{tabular}{cccc} \midrule
\parbox[c]{50pt}{Atom}&\parbox[c]{50pt}{$B_{\rm{fic}}$}&Groud state&\parbox[c]{50pt}{Angular\\momentum}\\
    \hline
\parbox[c]{50pt}{Li\\Na\\K\\Rb\\Cs}&\parbox[c]{50pt}{$5.4\times10^{-4}$\\$2.1\times10^{-2}$\\$0.21$\\$1$\\$3.6$\\}&$\left\}\parbox[c]{30pt}{{\hspace{2.5pt}}\\{\hspace{2.5pt}}\\{\hspace{2.5pt}}\\$^2S_{1/2}$\\{\hspace{2.5pt}}\\{\hspace{2.5pt}}\\}\right.$&$\left\}\parbox[c]{40pt}{{\hspace{2.5pt}}\\{\hspace{2.5pt}}\\electron\\spin\\{\hspace{2.5pt}}\\}\right.$\\
$\rm{^{171}Yb(I=1/2)}$&$6.7\times10^{-6}$&$^1S_0$    &nuclear spin\\
Tl                    &$2.5$&$^2P_{1/2}$&\parbox[c]{60pt}{orbital\\+electron spin}\\ \midrule
  \end{tabular}
  \end{center}
\end{table}

\section{Summary}

In summary, we have proposed a new spin-manipulation method using a quasi-electrostatic field. In addition, we have experimentally demonstrated the method by observing the Rabi oscillation of the spin of $\rm{^{87}Rb}$ atoms using a circularly polarized $\rm{CO_2}$ laser. This method is naturally understood as a magnetic resonance by an oscillating fictitious magnetic field induced by a quasi-electrostatic field, and is of crucial importance for applications such as atomic spin qubit unitary operation in quantum computation in which atomic spins must be locally manipulated without heating. 

\section{Acknowledgement}
This work is supported by the Grant-in-Aid for the 21st Century COE "Center for Diversity and Universality in Physics" from the Ministry of Education, Culture, Sports, Science and Technology (MEXT) of Japan, Grant-in-Aid for Scientific Research of JSPS and Strategic Information and communications R\&D Promotion Programme (SCOPE-S). One of the authors (J. K.) acknowledges support from JSPS.

\bibliography{apssamp}

\begin{thebibliography}{99}
\bibitem{Footnote1} For example, the contributions from $6^2P_{3/2}$ and $6^2P_{1/2}$ is about 3000 times smaller than that from $5^2P_{1/2}$ and $5^2P_{1/2}$ states for Rb atom.
\end{thebibliography}

\end{document}